\documentclass[twocolumn,showpacs,preprintnumbers,amsmath,amssymb,floatfix]{revtex4-1}

\usepackage{graphicx}
\usepackage{dcolumn}
\usepackage{bm}
\usepackage{epsfig}
\usepackage{epstopdf}
\usepackage{subfigure}

\begin{document}

\title{Crossing on hyperbolic lattices\\ }

\author{Hang Gu}%
 \email{ghbright@umich.edu}
 \author{Robert M. Ziff}%
 \email{rziff@umich.edu}
\affiliation{%
Michigan Center for Theoretical Physics and Department of Chemical Engineering, University of Michigan, Ann
Arbor MI 48109-2136.}

\date{\today}%

\begin{abstract}

We divide the circular boundary of a hyperbolic lattice into four equal intervals, and study the probability
of a percolation crossing between an opposite pair,
as a function of the bond occupation probability $p$.  We consider the \{7,3\} (heptagonal),  enhanced or extended binary tree (EBT), the EBT-dual, and \{5,5\} (pentagonal) lattices.  We find that the crossing probability increases gradually from zero to one as $p$ increases from the lower $p_l$ to the upper $p_u$ critical values.  We find bounds and estimates for the values of $p_ l $ and $p_u$ for these lattices, and identify the self-duality point $p^*$ corresponding to where the crossing probability equals $1/2$.  Comparison is made with recent
numerical and theoretical results.
\end{abstract}
\maketitle

\section{Introduction} 

Hyperbolic lattices represent curved surfaces in a space that is effectively of infinite dimensions.
While long of interest to mathematicians 
\cite{Beltrami1868}, and even artists \cite{Ernst07}, such lattices have only relatively recently been studied
in statistical physics, where many problems
 \cite{SakaniwaShima09,WuHolme10,SaussetToninelliBiroliTarjus10,BaekMinnhagenShimaKim09,ShimaSakaniwa06b,MadrasWu05,Wu00,SwierczakGuttmann96,KrioukovPapadopoulosFragkiskosKitsakVahdatBoguna10,boguna10,BaekMakelaHarriMinnhagenKim11} including
percolation \cite{Wu97,BenjaminiSchramm01,Lalley01,BaekMinnhagenKim09,NogawaHasegawa09,BaekMinnhagenKim09b,NogawaHasegawa09b,MadrasWu10,BaekMinnhagenKim10,MinnhagenBaek10,BaekMinnhagen11,BaekMinnhagen11b,Czajkowski11,Thale11} have been examined.   Investigation has also been carried out on 
 closely-related hierarchical lattices 
\cite{AutoMoreiraHerrmannAndrade08,AndradeHerrmannAndradedaSilva05,BoettcherCookZiff09,BoettcherSinghZiff11}.  
The study of hyperbolic lattices helps in the understanding of how geometry affects
 the behavior of systems.  There are also 
physical systems that show negative curvature on the nanoscale \cite{ParkEtAl03}. 
Networks are not confined to a physical dimensionality and can 
show hyperbolic behavior, and the current strong interest in network physics is another motivation
to study these systems.

Connectivity in networks is described by percolation, which has been studied for a wide variety of 
systems for well over 50 years \cite{StaufferAharony94}.   For most systems in 
percolation, there
is typically a bond or site occupation probability $p$, such that when $p$ is less
than a threshold value $p_c$, all connected components are finite, while above $p_c$
there is an infinite network which connects every region of the system.

For hyperbolic lattices, however,  percolation shows a more complicated behavior, with two distinct transitions,
as described below.   Various numerical and theoretical methods have been developed
to study these transitions, and in particular to find the threshold values \cite{Wu97,BenjaminiSchramm01,Lalley01,BaekMinnhagenKim09,NogawaHasegawa09,BaekMinnhagenKim09b,NogawaHasegawa09b,MadrasWu10,BaekMinnhagenKim10,MinnhagenBaek10,BaekMinnhagen11,BaekMinnhagen11b,Czajkowski11,Thale11}.

In this paper, we investigate percolation on hyperbolic lattices by considering the crossing probability, a technique which has not 
been applied to this system before.  
The lattices we study are some that have been considered by others, so that a comparison of
the values for the transition points  can be made.   We also study (for the first time numerically) percolation
on the pentagonal lattice, which has some interesting self-dual features.   In general, the determination of the
transition points allows one to compare the different numerical methods and
to test theoretical predictions and bounds.   These threshold points will be useful for future
studies of the nature of the transition behavior, such as the determination of critical exponents.

\begin{figure}[htbp]
\centering
\includegraphics[width=2in]{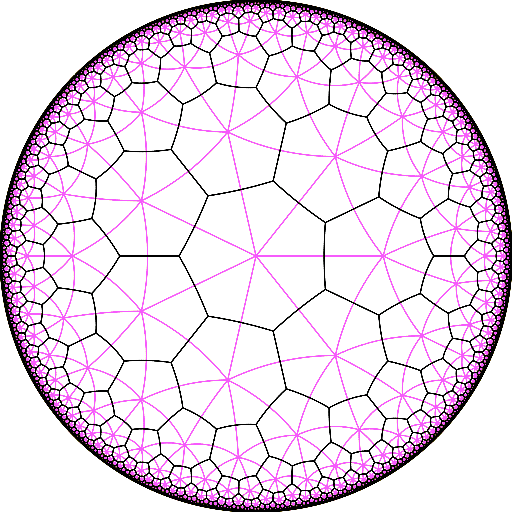}
\caption{(color online) Heptangonal lattice $\{7,3\}$ (black or dark)  and the dual lattice $\{3,7\}$  (magenta or light). }
\label{fig:hept}
\end{figure}

\begin{figure}[htbp]
\centering
\includegraphics[width=2in]{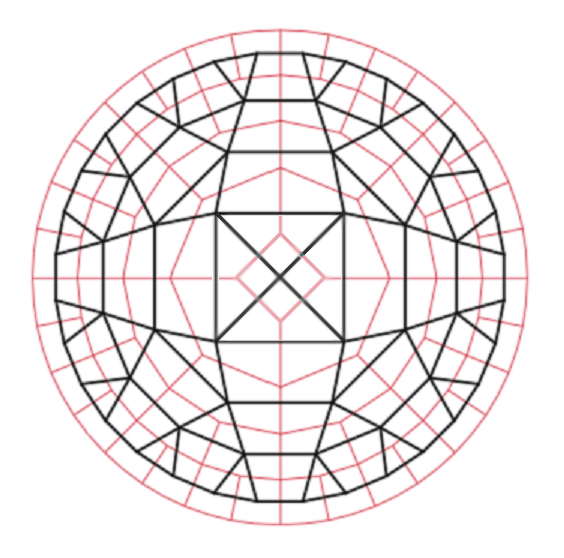}
\caption{(color online) The Enhanced Binary Tree (EBT) lattice (black or dark) and EBT-dual lattice (red or light).  The EBT is made by joining
four trees together.  We joined four bonds at the center rather than having two join there as in \cite{BaekMinnhagen11}.}
\label{fig:ebt}
\end{figure}

Common examples of hyperbolic lattices are those composed of identical polygons of $n$ sides, $m$ of which meet at a vertex.
These lattices can be characterized by the Schl\"afli symbol $\{n , m\}$, corresponding to the Gr\" unbaum-Shepard \cite{GrunbaumShephard87}
notation $(n^m)$.
Thus,  $\{6 , 3\}$ is a regular planar hexagonal (honeycomb) lattice, while $\{7 , 3\} $ is the heptagonal 
hyperbolic lattice shown in Fig.\ \ref{fig:hept}.  The dual
to the heptagonal lattice is the  $\{3 , 7\} $ lattice, also shown in Fig.\ \ref{fig:hept}.
The self-dual hyperbolic $\{5 , 5\} $ is shown in Fig. \ref{pentagonallattice}.

Recently, another type of hyperbolic lattice has been introduced, the Enhanced or Extended Binary Tree (EBT)
\cite{Wu97,NogawaHasegawa09}), which is made by
adding transverse bonds to the Bethe lattice.  The EBT, which is simpler to represent and code on a computer,
has been studied extensively for percolation \cite{BaekMinnhagenKim09b,NogawaHasegawa09,NogawaHasegawa09b,BaekMinnhagenKim10,MinnhagenBaek10,BaekMinnhagen11}.
The EBT and its dual are shown in Fig.\ \ref{fig:ebt}.

The general picture that has emerged for percolation on hyperbolic lattices \cite{Wu97,BenjaminiSchramm01,BaekMinnhagenKim09,NogawaHasegawa09}
is that there are two thresholds $p_ l $
and $p_u$, and the behavior is continuous between them.
For $0 < p < p_ l $, there are no ``infinite" (large) clusters connecting the
central area to the boundary sites,
for the intermediate region $p_ l  < p < p_u$ there are many infinite clusters touching the boundary,  and for $p_u < p < 1$,
there is exactly one infinite cluster.  These three regions persist in the limit that the system size goes
to infinity.  This behavior is in contrast to ordinary percolation, in which there
is no intermediate region (in an infinite system) and the crossing behavior is discontinuous.




Here we study percolation on hyperbolic lattices by examining a suitably defined crossing probability $R(p)$ as a function of the bond occupation probability $p$. 
The crossing probability is often studied in ordinary percolation to locate the
threshold and to investigate the critical scaling behavior \cite{ReynoldsStanleyKlein80,ZiffNewman02,HoviAharony96}.  
There, the crossing probability (the probability of a continuous path from one opposite side to another)
becomes steeper as the size of the system is increased, with a slope proportional to $L^{1/\nu}$, where $\nu$ is the correlation-length exponent,
equal to $4/3$ in two dimensions.  This behavior
defines the transition point uniquely in the limit of $L \to \infty$ \cite{StaufferAharony94}. 
Furthermore, when the system boundary is symmetric such as a perfect square, the crossing probability between opposite
sides is exactly $1/2$ 
(because the dual
lattice percolates if the original lattice does not) \cite{Cardy92,Ziff92}.  A crossing
probability of $1/2$ applies to a disk also, with
the circumference divided into four equal intervals.  In this paper, we set up a similar
 crossing problem for hyperbolic system by dividing the boundary of a finite system
into four equal-size intervals (or as equal as possible), and study the probability
of crossing between an opposite pair of these intervals.  We consider the heptagonal, EBT, EBT-dual, and pentagonal
hyperbolic lattices, and investigate how the resulting crossing
probability behaves with $p$.  We also discuss how the value of $p = p^*$ that corresponds
to a crossing probability of exactly $1/2$ relates to the two transition points $p_l$ and $p_u$. 

In the following sections we discuss the methods (II), the results (III), and the conclusions (IV).

\section{Method}

We begin by generating a hyperbolic lattice to a fixed number of generations or levels, so that the outside
is essentially circular as in Fig.\ \ref{fig:hept}.
Practically, it is only feasible to generate a relatively small number of levels (up to 10 -- 15) because of the exponential growth in the number of lattice sites with level.

For the heptagonal lattice with an open heptagon centered at the origin, 
we can derive an expression for the number of sites $N(l)$ as a function of level $l$ as follows:
Let $a_l$ equal the number of new sites which connect to the next generation,
and $b_l$ equal the number of new sites which connect to the previous generation.
Inspection of the Fig.\ \ref{fig:hept} shows that we have the relations
\begin{eqnarray}
a_l &=& 3 a_{l-1} + b_l \cr
b_l &=& a_{l-1}  
\end{eqnarray}
Thus, the total number of sites up to level $l$ is equal to $\sum_{l'=1}^l (a_l + b_l)$.
By means of generating function techniques, we find the explicit relation
\begin{equation}
N(l) = 7\left[\left(\frac{3 + \sqrt{5}}{2}\right)^ l  + \left(\frac{3 - \sqrt{5}}{2}\right)^ l - 2 \right]
\label{Nheptagonal}
\end{equation}
which yields $N( l ) = 7,$ 35, 112, 315, 847, 2240, 5887, 15\,435, 40\,432, 105\,875$\ldots$   for $ l  = 1, 2, \ldots, 10,\ldots $.
In Ref.\ \cite{BaekMinnhagenKim09}, the corresponding formula for $N(l)$ with
 a vertex rather than an open heptagon at the center of the system is given.  For large $ l $,
$N( l )$ grows exponentially as $\sim  7 [(3 + \sqrt{5})/2]^ l $.
These $N( l )$ are related to other mathematical quantities, such as the number of fixed points of period $ l $ in
iterations of Arnold's cat map at its hyperbolic fixed point, multiplied by 7  \cite{Arnold67}.

For the pentagonal lattice, we find
\begin{eqnarray}
a_l &=& 5 a_{l-1} + 3 b_l \cr
b_l &=& 3 a_{l-1} + 2 b_l  
\end{eqnarray}
which yields
\begin{equation}
N( l ) = \left[\left(\frac{7+ 3\sqrt{5}}{2}\right)^ l  + \left(\frac{7 - 3\sqrt{5}}{2}\right)^ l - 2 \right]
\label{Npentagonal}
\end{equation}
and equals 5, 45, 320, 2205, 15\,125, 103\,680, 710\,645, 4\,870\,845$\ldots$   for $ l  = 1, 2, \ldots, 8,\ldots $.
Here the $N( l )$  are related to the Fibonacci numbers $F(l) $ by $N( l ) = 5 \cdot F(2l)^2$.  


For the EBT, we consider a geometry with four trees meeting at the origin as shown in Fig.\ \ref{fig:ebt}, so
that it is easy to divide the boundary into four equal intervals.  The number of sites grows as
$N( l ) =  2^{ l + 2 }  - 3$.   For the EBT-dual, we have  $N( l ) = \cdot 2^ { l + 2 }  - 4$.

We applied the algorithm of \cite{NewmanZiff00,NewmanZiff01}
to find the crossing probability on these four lattices.  
This algorithm allows one to find an estimate of $R(p)$ for all values
of $p$ in a single sweep of the lattice; averaging over many sweeps
yields an accurate estimate of $R(p)$. 
The connections
between points in a cluster are represented by a tree structure, bonds are added 
one at a time,  and clusters are
joined together by means of a union-find operation.  The algorithm yields
the crossing probability $R_n$ as a function of the number of
occupied bonds $n$ added to the system, corresponding to a fixed-$n$ or  canonical ensemble.
To get the grand canonical result $R(p)$ corresponding to a fixed probability $p$,
one must convolve $R_n$ with the
binomial distribution:
\begin{equation}
R(p) = \sum_{n=0}^N \binom{N}{n} p^n (1 - p)^{N-n} R_n
\label{binomial}
\end{equation}
where $N$ is the total number of bonds in the system.
For large $N$, and the binomial distribution becomes very sharp, and for many problems it is not
necessary to carry out this convolution, but instead use just the
value at the maximum of the distribution $n = N p$, so that $R(p) \approx R_{N p}$.  However, for
smaller systems it is necessary to use this convolution to get accurate results.

\section{Results}

We carried out simulations for each of the four lattices,
recording $R(p)$ at $500$ values of $p$.  Below we describe the results for each lattice.


\subsection{Heptagonal \{7,3\} lattice }

\begin{figure}[htbp]
\centering
\includegraphics[width=3.4in]{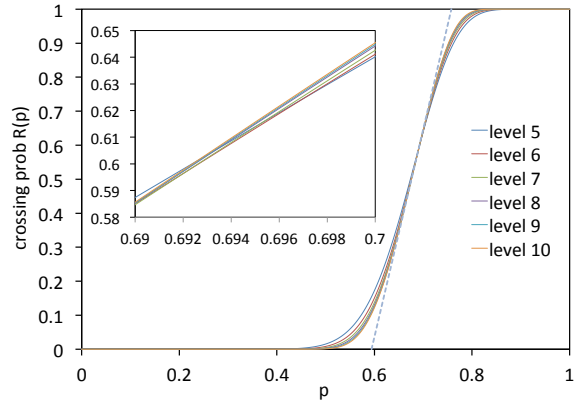}
\caption{(color online) Curves of crossing probability for the heptagonal lattice, convoluted with (\ref{binomial}), for various levels $ l $; the slope increases as $l$ increases.  The dashed line passes through the inflection point, and its
intercepts with the lines at $R = 0$ and $R = 1$ gives our estimates of $p_ l^B $ and $p_u^B$.  Inset: close-up near crossing point.}
\label{heptagonalcombined}
\end{figure}

\begin{figure}[htbp]
\centering
\includegraphics[width=3.4in]{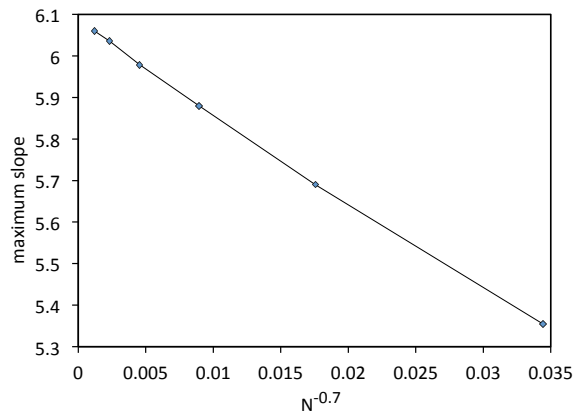}
\caption{Maximum of the slope of $R(p)$ {\it versus}\ $N^{-0.7}$ for the heptagonal lattice,
where $N$ is given by (\ref{Nheptagonal}).}
\label{maxslopeheptagonal}
\end{figure}

We considered the heptagonal $\{7,3\}$ lattice 
up to  level  $ l  = 10$ with $N( l ) = 105\,875$ total sites.
Fig.\ \ref{heptagonalcombined} shows the resulting $R(p)$ as a function of $p$ for
levels $5, \ldots, 10$.
We find a gradual transition of $R(p)$ from $0$ to $1$ as
the $p$ increases, as is typical for finite systems for ordinary percolation.  However,
here the width of the transition region is more spread out and, more significantly, the width 
limits to a non-zero value as $ l \to \infty $.  Equivalently the slope at the inflection point
limits to a finite value as $ l \to \infty $.
In Fig.\ \ref{maxslopeheptagonal} we plot the maximum
slope as a function of $N( l )^{-0.7}$ where $N( l )$ is given by (\ref{Nheptagonal}),
and find an extrapolation to a maximum value of $\approx 6.12$.
The exponent $-0.7$ was chosen empirically to get a fairly straight line; different choices
of the exponent do not change the intercept significantly and especially do not change
the conclusion that the slope limits to a finite value as $N \to \infty$.

\begin{figure}[htbp]
\centering
\includegraphics[width=3.4 in]{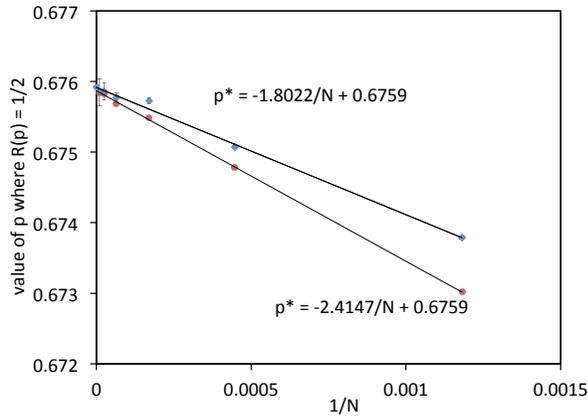}
\caption{(color online) Dual point $p^*(l) $ versus\ $1/N( l )$, where $N( l )$ is
the number of sites on the lattices of levels $ l  = 5, \ldots,10$ for the heptagonal lattice as given
by (\ref{Nheptagonal}).  We show results
for the raw (canonical) (upper points) and convoluted (grand canonical) (lower points) data; both extrapolate
to the same value, $p \approx 0.6759$, as $L \to \infty$.}
\label{heptagonatonehalf}
\end{figure}


The close-up in Fig.\ \ref{heptagonalcombined} shows that the curves do not quite cross at a 
single point, but the crossing point changes with $N$.  For a perfectly self-dual system 
in which the dual lattice is identical to the original lattice,
such as bond percolation on a square lattice and square boundary in ordinary percolation, 
the curves cross at a single point corresponding to $R = 1/2$ and $p= 1/2$, but because this
system is not self-dual, one would not expect the crossing to be at $(1/2,1/2)$ here.

We define the duality point $p^*(l)$ as the value of $p$ where $R(p) = 1/2$.
We call this the duality point because on a truly dual lattice
the crossing probability should also be $1/2$ (although below we see
that there are differences in the center that limit the extent that one can
make a completely self-dual system). 
We find  $p^*(\infty) \approx 0.6759$ by extrapolating to $l = \infty$
as shown in Fig.\ \ref{heptagonatonehalf}.  Here we observed a scaling
of order $1/N(l)$.

The transition points for the heptagonal lattice were found by Baek et al.\  
\cite{BaekMinnhagenKim09} to be $p_ l  \approx 0.53$ and $p_u \approx 0.72$, 
and on the dual lattice $\{3,7\}$ they found $p_ l  \approx 0.20$ and $p_u \approx 0.37$.
These four values are evidently not completely consistent because one should have, for any lattice and its dual,
\begin{eqnarray}
p_ l  + p_u^\mathrm{dual} &=& 1 \cr
p_u + p_ l ^\mathrm{dual} &=& 1
\label{dualequations}
\end{eqnarray}
 
One would expect that for $p > p_u$, $R(p) = 1$, and for $p < p_l$, $R(p) = 0$.  However,
how $R(p)$  approaches those values from the region $p_l < p < p_u$ is not clear.  It appears
from our data the approach is tangential (with slope zero), and therefore it is rather
hard to identify the transition points accurately.  We can, however, find bounds to that
behavior by drawing a tangent line from the inflection point.
Drawing a line through the inflection point in 
Fig.\ \ref{heptagonalcombined} with the maximum slope $\approx 6.12$, and the intercepts
for $R(p) = 0$ and $R(p) = 1$ give us the rather crude bounds  $p_ l  < p_ l^B =  0.594$ and $p_u > p_u^B = 0.758$.

\begin{figure}[htbp]
\centering
\includegraphics[width=3.4in]{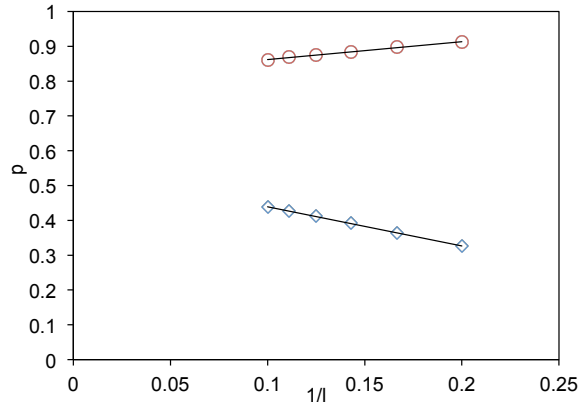}
\caption{Values of $p$ where $R(p) = 1  - 10^{-5}$ (upper data points) and $R(p) = 10^{-5}$ (lower data points),
plotted as a function $1/l$, for the \{7,3\} heptagonal lattice.  The linear extrapolation
to $l \to \infty$ gives our estimates $p_u^e = 0.810$ and $p_l^e = 0.551$.  Extrapolations for the other
lattices show similar linear behavior and the values for $p_l^e$ and $p_u^e$ are given in Table \ref{table1}.}
\label{heptagonalestimates}
\end{figure}

A more precise method to get bounds or estimates for the transition point 
is to look at the values of $p$
where $R(p) = \epsilon$ and $R(p) = 1 - \epsilon$, where we chose  $\epsilon = 10^{-5}$,
and then extrapolating
to $L = \infty$.  Fig.\ \ref{heptagonalestimates} shows that these estimates appear to scale as 
$1/l$, and extrapolating the points to $l \to \infty$ gives the values of 
$p_l^e$ and $p_u^e$ listed in Table \ref{table1}. 
In principle, the smaller value of $\epsilon$ the better, but noise in the data and precision of the numbers in our output files limited
how small we could make $\epsilon$.  The fixed number of points that we recorded (500) also limited the final precision
of the thresholds.
Varying the value of $\epsilon$ by an order of magnitude in each direction and finding the extrapolated
thresholds, we estimate that the overall error in our threshold estimates
is about $\pm 0.01$.

\subsection{EBT and EBT-dual lattices}

\begin{figure}[htbp]
\centering
\includegraphics[width=3.4 in]{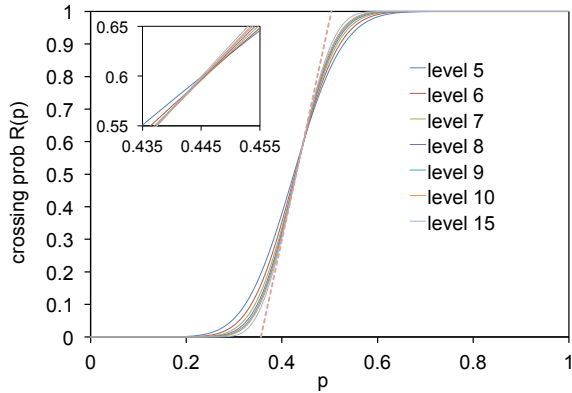}
\caption{(color online) The crossing probability $R$ as a function of $p$ for the EBT lattice, for systems of 5 -- 15 levels.
The dashed line passes through the inflection point.}
\label{EBTcrossing}
\end{figure}

\begin{figure}[htbp]
\centering
\includegraphics[width=3.4 in]{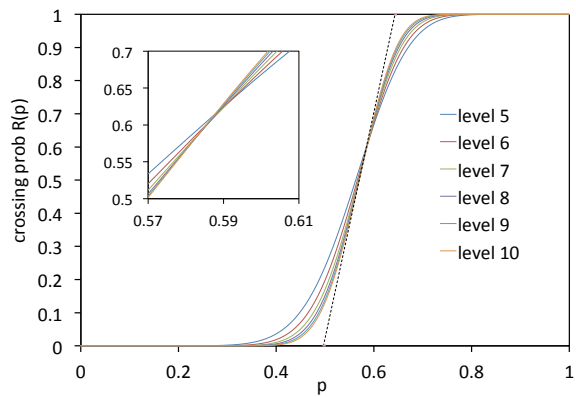}
\caption{(color online) The crossing probability $R$ as a function of $p$ for the EBT-dual lattice, for systems of 5 -- 10 levels.
The dashed line passes through the inflection point.}
\label{EBTdualcrossinginset}
\end{figure}

We simulated the EBT lattice to the level of 15, and the EBT-dual lattice to the level of 10.  Figs.~\ref{EBTcrossing} and \ref{EBTdualcrossinginset} show the resulting crossing probability distribution for these two lattices.  For the EBT, the maximum slope converges to $\approx 6.79$. Its duality point is at  $p^* \approx 0.4299$, yielding the bounds $p_ l^B  \approx 0.356$ and $p_u^B \approx 0.503$.  The EBT-dual's crossing probability distribution curve also converges to a maximum  slope $\approx 6.83$, the duality point of which is at  $p^* \approx 0.5698$, yielding the bounds $p_ l^B = 0.497$ and $p_u^B =  0.643$.   
These bounds satisfy the
expected duality (\ref{dualequations})  within errors.  The estimates are also found to the scale as $1/l$ and the resulting values $p_l^e$ and $p_u^e$ given in Table \ref{table1}.
These estimates do not satisfy the duality relations (\ref{dualequations}) very precisely, reflecting rather large error bars in their values.

\subsection{Pentagonal lattice}
\begin{figure}[htbp]
\centering
\includegraphics[width=1.8in]{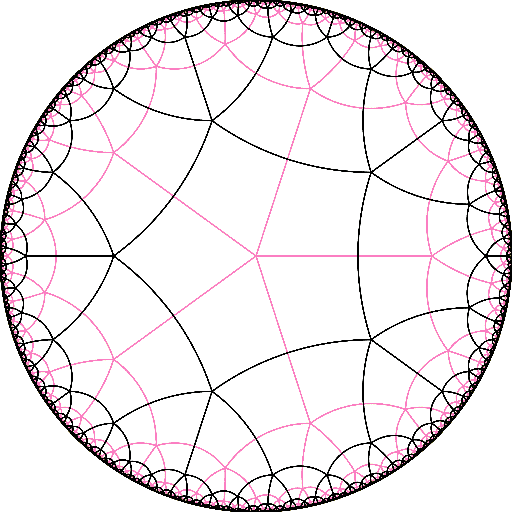}
\caption{(color online) Pentagonal (black or dark) and dual pentagonal (red or light), both $\{5,5\}$.}
\label{pentagonallattice}
\end{figure}

\begin{figure}[htbp]
\centering
\includegraphics[width=3.4 in]{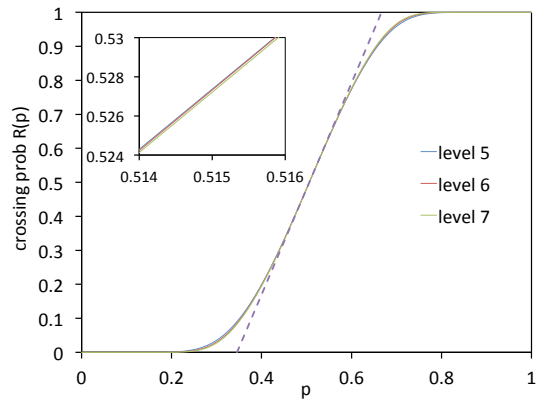}
\caption{(color online) The crossing probability $R$ as a function of $p$ for the \{5,5\} or pentagonal lattice, for systems of 5, 6, and 7 levels.  The curves are nearly indistinguishable on this plot.  The dashed line passes through the inflection point.}
\label{pentagonalcrossing}
\end{figure}

We also considered the pentagonal $\{5,5\}$ lattice, which is shown in Fig.\ \ref{pentagonallattice}. This lattice is interesting because it is self-dual in an infinite system.  For the systems of a finite number of levels, it is not precisely self-dual because the center is different: on what we call the pentagonal lattice, there is a pentagon at the center, while for the pentagonal-dual, there is a vertex at the center  (see Fig~\ref{pentagonalcrossing}). 
We find that $p^* = 0.506 \pm 0.001$, so it is not exactly at $0.5$ as one might expect from duality.  Evidently, the central region plays an important role and a significant fraction of the percolating clusters connecting opposite sides pass through it, making the pentagonal and pentagonal-dual lattices slightly different with respect to the crossing problem we consider.

The slope of its crossing probability curve converges to a maximum value $\approx 3.12$, with bounds  $p_ l^B \approx 0.346$ and $p_u^B \approx 0.666$, which indicates the distribution is nearly symmetric.  

Recently, Delfosse and Z\'emor \cite{DelfosseZemor10} have shown that, for any self-dual hyperbolic lattice $\{m,m\}$, $ 1/(m-1) \le p_l \le 2/m$, so that for $m = 5$,  $1/4 \le p_l \le 2/5$.  Our bounds $p_l^B$ and $p_u^B$ fall well within these values, and our estimates $p_l^e$ and $p_u^e$ are close to the bound $1/4$ and (by duality) $3/4$, respectively.  This bound follows
from approximating the hyperbolic lattice as a tree (Bethe lattice) of coordination number 5.

\begin{table}[htbp]
   \centering
   \begin{tabular}{|c|c|c|c|c|c|c|}
      \toprule
            Lattice    &  $p^*$& max.\ slope &$p_ l^B $ & $p_u^B$ & $p_ l^e $ & $p_u^e$ \\
      \hline
       \{7,3\}   &0.6759 & 6.12  & 0.594 & 0.758 & 0.551 & 0.810\\
       EBT   & 0.4299 & 6.79  & 0.356 & 0.503 & 0.306 & 0.564\\
       EBT-dual   & 0.5698 & 6.83  & 0.497 & 0.643 & 0.452 & 0.699\\
       \{5,5\}   & 0.506 & 3.12  & 0.346 & 0.666 & 0.263 & 0.749 \\
                \hline
   \end{tabular}
   \caption{Values of the dual point $p^*$ where $R(p) = 1/2$, the slope at that point, our bounds $p_ l ^B$ and $p_u^B$ for the various lattices we studied, such that $p_l < p_l^B$, and $p_u > p_u^B$, and our extrapolated estimates of the transition points $p_ l ^e$ and $p_u^e$ .  In general, the numbers are expected to be accurate to about $\pm 1$ in the last digit shown, except the estimates $p_ l^e $ and $p_u^e$, which are expected to be accurate to about $\pm 10$ in the last digits.}
   \label{table1}
\end{table}

\begin{table}[htbp]
   \centering
   \begin{tabular}{|c|c|c|c|}
      \toprule
            Lattice    &  $p_ l $ & $p_u$ & Ref.\\
      \hline
       \{7,3\}   & $0.53$& $0.72$ &\cite{BaekMinnhagenKim09}   \\
       \{3,7\}   & $0.20$ & $0.37$  & \cite{BaekMinnhagenKim09}  \\
       EBT   & 0.304(1) & 0.564(1) & \cite{NogawaHasegawa09}  \\
        "  &  & 0.48 & \cite{BaekMinnhagenKim09}  \\
         "  & $(\sqrt{13}-3)/2 \approx 0.3028$& 0.5 & \cite{MinnhagenBaek10}  \\
       EBT-dual   & 0.436(1) & 0.696(1)& \cite{NogawaHasegawa09} \\
       \{5,5\}   & $0.25\le p_ l\le 0.4$ &    &\cite{DelfosseZemor10} \\
         \hline
   \end{tabular}
   \caption{Previous values of the transition points.  }
   \label{table2}
\end{table}

\section{Conclusions}

In summary, we have the following results and conclusions:

(i)  The crossing probability approaches a continuous S-shaped curve with a finite maximum slope at the
inflection point as $l \to \infty$.

(ii)  By drawing a tangent line through the inflection point and finding its intercept with $R(p)=0$ and $R(p)=1$, we find the bounds 
$p_l^B$ and $p_u^B$ for the transition points $p_l$ and $p_u$ listed in Table \ref{table1}.  
Also, by extrapolating where $R(p) = \epsilon$ and $R(p) = 1 - \epsilon$ to $L \to \infty$, we find the estimates $p_l^e$ and $p_u^e$,
also listed in Table \ref{table1}.
 In comparison,  previously measured and predicted values of $p_l$ and $p_u$ (determined through other methods) are listed in Table \ref{table2}.
 
(iii) For the \{7,3\} lattice, the reported value $p_u = 0.72$ \cite{BaekMinnhagenKim09} is inconsistent with our lower bound $p_u^B  = 0.758$ and estimate $0.810$.
However, those authors' value for $p_l = 0.20$ on the dual lattice \{3,7\} is consistent with this bound, by (\ref{dualequations}).

(iv)   For the EBT lattice, our bound $p_u^B = 0.503$ and especially our estimate $p_u^e = 0.564$ are inconsistent with the prediction $p_u = 1/2$ \cite{MinnhagenBaek10}.  Our results for $p_u$ are however consistent with the measurement of $p_u$  by \cite{NogawaHasegawa09}.  For $p_l$, our estimate $0.306$ is in substantial agreement with the results of both Refs.\ \cite{BaekMinnhagenKim09} and \cite{NogawaHasegawa09}.

(v)  For the EBT-dual lattice, our bounds and estimates for the transition points agree with those of \cite{NogawaHasegawa09} within expected errors.

(vi) For the \{5,5\} lattice, we report measurements of the thresholds
for the first time, and our estimates
are close to the theoretical bounds $p_l = 1/4$ and $p_u = 3/4$, which follow
by assuming  a 
tree structure  \cite{DelfosseZemor10}.

(vii)  We determine the point $p^*$ where $R(p^*) = 1/2$ for all four lattices we consider,
and find that the behavior of $R(p)$ is nearly symmetric about that point.
For  the \{5,5\} lattice, $p^* \approx 0.506$, slightly larger than the value
$0.5$  one might expect from self-duality.  We believe the deviation from $0.5$ is
due to the non-equivalent configurations at the center for the lattice and its dual, 
implying that the two finite systems are not exactly dual to each other.

(viii) We have found a variety of finite-size scaling relations, as given in 
Figs.\ \ref{maxslopeheptagonal}, \ref{heptagonatonehalf}, and \ref{heptagonalestimates}.
Another scaling relation that exists in the literature is that of
Nogawa and Hasegawa \cite{NogawaHasegawa09}, who find that the mass of the root cluster
in the EBT scales as $N^{\psi(p)}$ in the intermediate region, 
where $\psi$ is a function of $p$.
Clearly it would be desirable to have a general scaling
theory that combines all of these finite-size scaling relations.   
This is an interesting problem for future study. 

(ix)  Another area for future study is site percolation on hyperbolic
lattices.  For site percolation on fully triangulated lattices in ordinary two-dimensional
space, $p_c = 1/2$.   For fully triangulated 
hyperbolic lattices, such as the $\{3,7\}$ lattice, we guess that the behavior of $R(p)$ will be
precisely symmetric about $p = 1/2$, because of its self-matching property.
Also, because bond percolation on a given lattice is equivalent to site percolation
on its covering lattice (or line graph), the results here for bond percolation on 
the  \{7,3\} lattice can be mapped to site percolation on its covering lattice,
which is an interesting kind of hyperbolic kagom\'e lattice.  Covering
lattices of the other lattices we considered here contain crossing bonds.

\emph{Note added in proof:}  Just recently, Baek \cite{Baek12} argued that the conjectured result $p_u = 0.5$ for the EBT in Ref. [23] should be replaced by a lower bound of about  0.55, which is consistent with our results here.

  
\section{Acknowledgments}
Discussions and correspondence with Stefan Boettcher, Tomoaki Nogawa, Seung Ki Baek and Nicolas Delfosse are gratefully acknowledged.  
\bibliographystyle{elsarticle-num}

\bibliography{HyperbolicLatticesPREFinal.bib}

\vfill\eject

{\bf Notes added after publication:}
This paper has been published in Physical Review E {\bf 85}, 051141 (May 29, 2012).  We add  the following
update and correction:

1.   Shortly after our paper was published, Delfosse and Z\'emor [45] posted a paper
in which  the rigorous upper bound for $p_l$ for the \{5,5\} lattice is
reduced from 0.40 to 0.38, closer to our approximate bound $p_l^B = 0.346$.   They also give upper bounds for self-dual lattices $ \{m,m\} $ for larger values of
$m$.

2. A correction in the printed version:  In section III A, third paragraph, the second sentence should
read: ``We call this the duality point because on a truly
dual lattice the \emph{crossing} probability should also be 1/2\ldots."  We have changed ``occupation"
to ``crossing" in the present version.

\bigskip

[45]  Nicolas Delfosse and Gilles Z\'emor, ``Upper Bounds on the Rate of Low Density Stabilizer Codes for the Quantum Erasure Channel,"  arXiv:1205.7036 (May 31, 2012).

\end{document}